# Magnetization dynamics of weak stripe domains in Fe-N thin films: a multi-technique complementary approach.

I S Camara,[1] S Tacchi,[2] L- C Garnier,[1,3] M Eddrief,[1] F Fortuna,[5,6] G Carlotti,[4] and M Marangolo[1]

[1] Sorbonne Universités, UPMC Universités Paris 06, CNRS,

Institut des Nanosciences de Paris, 4 place Jussieu, 75252 Paris, France

[2] Istituto Officina dei Materiali del CNR (CNR-IOM), Sede Secondaria di Perugia,

c/o Dipartimento di Fisica e Geologia, Università di Perugia, I-06123 Perugia, Italy

[3] Université Versailles St-Quentin, LISV, Bâtiment Boucher,

Pôle scientifique et technologique de Vélizy, 10-12 avenue de l'Europe, F-78140 Vélizy, France

[4] Dipartimento di Fisica e Geologia, Università di Perugia, Via Pascoli, I-06123, Perugia, Italy

[5] Univ Paris-Sud, CSNSM, UMR 8609, Bâtiments 104 et 108, F-91405 Orsay, France

[6] CNRS, IN2P3, F-91405 Orsay, France

E-mail: tacchi@iom.cnr.it

**Abstract**

The resonant eigenmodes of a nitrogen-implanted iron α'-FeN characterized by weak stripe domains are investigated by Brillouin light scattering and broadband ferromagnetic resonance experiments, assisted by micromagnetic simulations. The spectrum of the dynamic eigenmodes in the presence of the weak stripes is very rich and two different families of modes can be selectively detected using different techniques or different experimental configurations. Attention is paid to the evolution of the mode frequencies and spatial profiles under the application of an external magnetic field, of variable intensity, in the direction parallel or transverse to the stripes. The different evolution of the modes with the external magnetic field is accompanied by a distinctive spatial localization in specific regions, such as the closure domains at the surface of the stripes and the bulk domains localized in the inner part of the stripes. The complementarity of BLS and FMR techniques, based on different selection rules, is found to be a fruitful tool for the study of the wealth of localized magnetic excitations generally found in nanostructures.

# 1. Introduction

In the continuous search for new materials with engineered properties, an important role is played by ferromagnetic films that develop magnetic stripe domains thanks to the presence of a strong perpendicular magnetic anisotropy (PMA). The occurrence of stripes can appreciably modify the value of the coercive field and of the microwave permeability, so that understanding the physics of stripe domains could facilitate the engineering of microwave devices made with PMA materials. Actually, magnetic stripe domains have been extensively studied [1,2,3,4] since their effects were firstly observed in Permalloy thin films by Prosen et al. [5]. These authors noted that the direction of the easy magnetization in Permalloy films could be selected by the application of a sufficiently large magnetic field along this direction and called this new property rotatable anisotropy. Three years later, Saito et al.[6] showed that this rotatable anisotropy is due to the appearance of magnetic stripe domains (SD) whose axis is parallel to the last saturation field. It is now agreed that SDs appear in ferromagnetic films with perpendicular anisotropy $K_u$. This latter can originate from magnetocrystalline and/or stress anisotropy[7,8,9]. Other parameters such as the film thickness $t_f$, the saturation magnetization $M_s$ and the exchange stiffness $A$ also play a significant role on the nucleation and characteristics of SDs [10]. For materials with strong perpendicular anisotropy in comparison with the demagnetizing energy $\mu_0 M_s^2/2$, i.e. where the quality factor $Q=2K_u/\mu_0 M_s^2$ is larger than 1, it was shown that "strong stripe" domains appear. They are characterized by large domains uniformly magnetized perpendicularly to the film plane and separated by infinitely thin one-dimentional Bloch domain walls [10,11,12]. For materials with low perpendicular anisotropy ($Q < 1$), instead, two situations can occur depending on the film thickness[10]: below some critical thickness $t_c$ the magnetization lies in the plane of the thin film, while above $t_c$ the magnetization develops a periodic out-of-plane component of sine-like profile [6,10,13] whose half period $W$ is comparable to the film thickness $t_f$. The latter regime is called "weak-stripe" structure and is characterized by a significant inhomogeneity of the magnetization across the thickness of the film[10]. Unlike the strong stripe regime, in the weak stripe regime the Bloch wall height $t_B$ can be significantly reduced compared to the film thickness $t_f$ ($t_B/t_f < 0.6$), [14] because of the formation of Néel-type flux closure caps at the film surface.

The presence of weak stripe domain structures has been also reported in FeGa[15,16,17], FePd[13], NiPd[18], and Co/Fe multilayers[19]. Lately, a weak stripe domain structure was observed by some of us in nitrogen-implanted iron α'-FeN films [20,21] Here, stripe domains with a period of about 130 nm were found, at remanence, along the last saturating magnetic field direction.

Several studies have been devoted to understanding the physics of the above stripe domains focusing on the nucleation conditions of the stripes, on the magnetization distribution and on the in-plane magnetization reversal mechanisms. Among these studies, we recall here theoretical models [10,22,23] and micromagnetic simulations [19,24,25,26,27], as well as experimental studies carried on by both quasistatic techniques, like magnetic force microscopy[4,15] (MFM), x-ray magnetic microscopy[28] or vibrating sample magnetometer[17,20] (VSM), and dynamic ones, such as Brillouin light scattering[16] (BLS) and ferromagnetic resonance [29,30,31] (FMR). Very recently, spin-wave propagation in a Co/Pd film with stripe domains has also been investigated,[32] showing that the spatial periodicity of the magnetization serves as a periodic potential for spin waves propagation, as it happens in artificial magnonic crystals. As a result, the number of modes increases in the BLS spectra, and their dispersion depends on the angle of propagation with respect to the direction of the stripes.

In this paper we present the results of a thorough study of the dynamic behavior of a 78-nm thick, nitrogen-implanted iron α'-FeN film, focusing our attention on the evolution of the resonant eigenmodes of the weak stripe domains on an external magnetic field, of variable intensity, applied in the film plane, either parallel or transverse with respect to the stripes direction. Once the basic information on the stripes structure and its evolution with the applied field is achieved by VSM and MFM, the experimental analysis of the resonant excitations will be performed comparing the results of two complementary experimental techniques, namely coplanar waveguide based broadband FMR (CPW-FMR) and BLS, with those of advanced micromagnetic simulations. The aim of this investigation is to rationalize the characteristics of the dynamical eigenmodes, correlating them to the underlying stripe domain structure as a function of the intensity of an external magnetic field. We will show that the spectrum of the dynamic eigenmodes in the presence of the weak stripes is very rich and that different families of modes can be selectively detected using different techniques or different experimental configurations. The different evolution of the modes frequencies with the intensity and the direction of the external magnetic field is accompanied by spatial localization in specific regions of the stripe domains. For instance, when the external field is reduced from saturation and the stripe domains appear, the perpendicular standing modes of the saturated films evolve assuming a preferential localization at the surface of the stripes. In addition, other modes, that are not present when the film magnetization is uniform, appear and localize in the bulk domains (inner part of the stripes). Instead, if an increasing external magnetic field is applied in-plane, but orthogonal to the stripe direction, the evolution of the modes is more complicated and a discontinuous transition is observed at the field value that corresponds to the sudden rotation of the stripe pattern.

## 2. Experiment and micromagnetic modeling

The sample has been prepared by ion implantation of nitrogen molecular ions $N^+_2$ on an α-Fe film with a thickness of 78 nm, following the procedure described in Ref.20. The iron film has been de-posited by molecular beam epitaxy (MBE) on ZnSe-buffered GaAs(001) substrate. Reflection high-energy electron diffraction (RHEED) measurements revealed that the [110] axis of α-Fe, ZnSe and GaAs were parallel to each other, as well as the [100] axis. The film was protected against oxidation by an 8-nm-thick gold capping layer. Ion implantation was performed at room temperature with nitrogen molecular ions $N^+_2$ accelerated to 40 keV with a fluence of $5.3 \times 10^{16}$ ions/cm$^2$. The ion current density was kept below 3μA/cm$^2$ and the targets were not cooled during the implantation. The crystal structure of the sample was studied by means of x-ray diffraction measurements in the out-of-plane direction using monochromatized Cu Kα radiation in a high-resolution SmartLab diffractometer. The formation of a body-centered tetragonal N-martensite whose c-axis is perpendicular to the film plane and c-parameter is close to that of α'-Fe$_8$N, has been observed. MFM measurements were performed using a Bruker Dimension AFM operating in the phase-detection mode. The in-plane hysteresis loops have been measured at room temperature by the VSM technique using a PPMS (Quantum Design). Additional VSM measurements have been done to study the stripes rotation.

CPW-FMR measurements were carried out placing the sample on top of a broadband channelized coplanar waveguide transducer with the magnetic layer facing it. The coplanar waveguide had a central track of width 300 μm. A microwave vector network analyzer (Rohde & Schwarz) was used to feed the waveguide with microwaves and to measure its transmission coefficient $S_{21}$ as a function of the microwave frequency f = 1-25 GHz and of external magnetic field μ$_0$H in the range between 400 mT and -400 mT. This is converted into a dimensionless quantity U = ($S_{21}$ - $S^{ref}_{21}$)/$S^{ref}_{21}$ where $S^{ref}_{21}$ is the value of the transmission coefficient measured for a reference field μ$_0$H$^{ref}$.

BLS measurements were performed in the backscattering configuration, focusing about 200 mW of monochromatic light (532 nm wavelength) onto the sample surface. The backscattered light was frequency analysed by a Sandercock-type 3+3-pass tandem Fabry-Perot interferometer.[33] The external magnetic field was applied parallel to the film surface, while the incidence angle was set to zero, i.e. the direction of the incoming light was normal to the sample plane

Micromagnetic simulations have been performed using the graphic processing unit (GPU) accelerated software Mumax3.[34] The saturation magnetization and the damping were respectively set to M$_s$ = 1700 kA/m, the value obtained from the VSM measurements, and α=0.01, in the range of values recently reported by T. Hwang et. al[35]. The exchange constant A and the anisotropy K$_u$ were estimated from the nucleation field μ$_0$H$_s$ = 350 mT, the stripes period P = 100 nm, the film thickness $t_f$ = 78 nm and the saturation magnetization $M_s$ by applying the procedure developed by Asti et al.[23] Their method is based on the calculation of the micromagnetic parameters A, $K_u$ that are solutions of the static equilibrium equation M×$H_{eff}$ = 0 linearized at the nucleation field $H_s$ by assuming small out-of-plane components of the magnetization. In the static equilibrium equation, M is the local magnetization vector and $H_{eff}$ is the local total effective field, obtained as the sum of exchange, anisotropy, magnetostatic and external in-plane fields. For our FeN sample, the exchange stiffness constant and the out-of-plane uniaxial anisotropy were estimated to be A = 21.6 pJ/m and $K_u$ = 5.2×$10^5$ J/$m^3$, respectively, according to Ref.21. Note that the estimated exchange constant of FeN is similar to the value A = 20 pJ/m of iron bulk[36] and that the calculated anisotropy is similar to the value ~ 5×$10^5$ J/$m^3$ previously reported by N. Ji et al.[37] The total simulated area has dimensions of 2000×2000×78 $nm^3$, and was discretized into cells having dimensions of 3.1×3.1×2.4 $nm^3$.

For the simulations of the magnetization dynamics, the excitation was realized by applying an exponential pulse, $\vec{h}(t) = \vec{h_0}exp(-t/\tau_{pulse})$ with amplitude of $h_0$= 60 mT and duration of $\tau_{pulse}$ = 100 fs. Different spatial profiles and directions of the linear polarization of the above exciting field were exploited. In the simulations aimed at reproducing the BLS results, we used an excitation pulse polarized perpendicularly to the surface of the film, having a maximum amplitude of 60 mT, and resulting from the superposition of several profiles with either even or odd nodal plane through the film thickness (see Fig.S1(a) of supplemental material). In addition, to mimic the limited penetration depth of light into the FeN film, it has been probed the dynamics of the component of the magnetization perpendicular to the film surface ($m_z$) down to 7.3 nm. In order to reproduce the FMR results, instead, the micromagnetic simulations have been carried out using an excitation pulse with a uniform spatial profile across the thickness and probing the dynamic magnetization through the whole film thickness. Since in an FMR experiment the power absorbed by the sample is given by $P_{abs} = \langle \vec{h} \cdot d\vec{m}/dt \rangle$ where $\vec{m} = \vec{M}/M_S$, we monitor the time dependence of the component of the magnetization $M_h(t) = M_h^0 + m_H(t)$ along the direction of the pulse field, where $M_h^0$ is the static magnetization. After the excitation by the field pulse, we recorded the magnetization dynamics of each discretized cell for a time length of 20 ns, sampled at a rate of 10 ps. A discrete Fourier transform of the time domain magnetization allowed us to obtain the spectrum of the probed magnetization component in the frequency domain. with a resolution of 50 MHz.

## 3. Results and discussion

### 3.1 MFM measurements

MFM measurements have been exploited to image the configuration of magnetic stripes. On reducing the intensity of the external field, applied in the sample plane, below about 350 mT, a well-defined stripe domains pattern, aligned along the field direction is observed (Fig.1(a)). From the fast Fourier transform of the images, we found that the period of the pattern is about 100 nm, as it can be seen in Fig.1(b) Moreover, MFM measurements have been performed to study the stripes rotation. The sample was initially saturated applying a field $\mu_0 H_{sat}$=0.4 T external magnetic field along the [100] direction to generate stripe domains aligned along that direction. Then a magnetic field $\mu_0 H_{trans}$ of increasing intensity was applied along the [010] axis in the direction perpendicular to the stripes axis. MFM images were recorded at remanence after $\mu_0 H_{trans}$ is turned off. The whole stripe structure was found to remain unperturbed up to about $\mu_0 H_{trans}$ = 100 mT, then the stripes coherently rotate with an angle $\theta$=60° towards the field direction. On increasing the transverse field the rotation angle $\theta$ increases until a complete reorientation of the stripes is fulfilled above 200 mT. During the stripes rotation, the period of the pattern was found to remain almost constant to about 100 nm.[20]

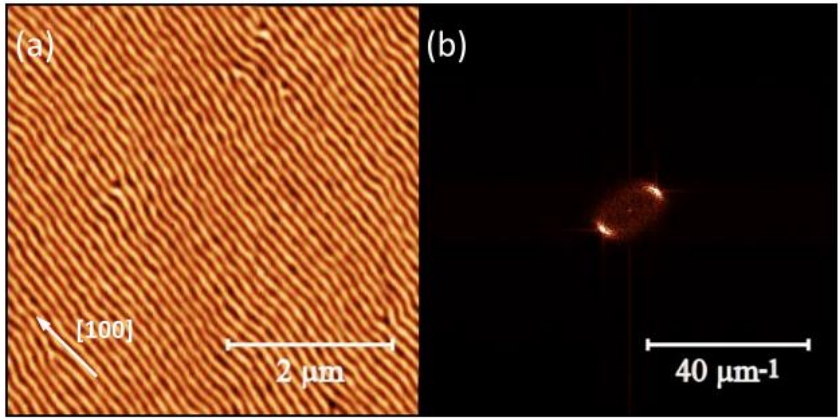

**Fig.1** (a) MFM images taken at remanence after in-plane saturation along the [100] direction. (b) Two-dimensional fast Fourier transform of the MFM image reported in the panel (a).

*3.2 VSM measurements and static micromagnetic simulations*

Longitudinal hysteresis loops were measured at room temperature by applying a field $\mu_0H_{long}$ along the [100] and the [110] directions by VSM. While pure (001)-textured α-Fe film is characterized by an in-plane biaxial anisotropy, with the easy axis along [100] direction, an isotropic in-plane magnetic behavior has been observed for the α-FeN film. A typical hysteresis loop, measured along the [110] axis, is displayed in Fig.2(a) as a dashed black line. Starting from positive saturation a linear M(H) dependence starts at $\mu_0H_{long}$ close to 350 mT, suggesting the formation of stripe-like domains parallel to the applied magnetic field, in good agreement with MFM results. The saturation magnetization is $M_s$ = 1700 kA/m, while the coercive field is 25 mT. The simulated hysteresis curve is reported in Fig.2(a) as a continuous red line and as it can be seen it exhibits a good agreement with the experimental one. Fig.2(c) shows a cross-section (perpendicular to the stripes direction) of the magnetization distribution calculated at remanence. One can observe basic domains (region e), where the magnetization points alternately up and down with respect to the film plane with a slight tilt near the surface. Between two adjacent basic domains, a Bloch-type transition is observed inside the thin film (region f), while a flux-closure domain of Néel-type transition is observed at the film surfaces (region g). The average volume fractions $V_e$=73% and $V_g$=25% of basic and flux-closure domains, respectively, are estimated from the average of the normalized magnetization components at remanence. This high relative volume fraction $V_g$ of the flux-closure domains is characteristic of the weak stripe domain regime[22] and it is in accordance with the values of the quality factor $Q = 2K_u/\mu_0M^2_s$=0.29 and the reduced thickness $t_r = t_f/\sqrt{A/K_u}$ [17]. VSM measurements have been also carried out during the stripes rotation. First, the stripe domains were prepared along the [110] axis. Then a magnetic field $\mu_0H_{trans}$ of increasing intensity was applied along the [1-10] axis and the component of magnetization $M_{trans}$ along the $H_{trans}$ direction was recorded. As shown in Fig.1(b) (dashed lines), $M_{trans}$ increases linearly with the transverse field $\mu_0H_{trans}$ between 0 and 350 mT, reaching a saturated state above 350mT. A considerable reduction of the slope is observed for fields larger than 100 mT, suggesting the onset of a new regime due to the rotation of the stripes in agreement with the MFM measurements. In order to explain the behavior of $M_{trans}$, micromagnetic simulations have been also performed applying a magnetic field perpendicular to the stripes domains. When a small magnetic field is applied, the basic domains, which are out-of-plane magnetized, remain nearly unchanged, while the flux closure domains with in-plane magnetization component parallel (antiparallel) to the field direction, expand (shrink), as it can be seen in Fig.2 (d). This phenomenon has been extensively described in Ref. 17. Therefore, the component of the magnetization parallel to the external field linearly increases with $H_{trans}$, in agreement with the VSM meas-urements (red curve in Fig.2(b)) and results reported in Ref. 20. Above 200 mT, the external magnetic field makes the stripes rotate towards the applied field direction, causing a reduction of the $M_{trans}$ slope. Note that the higher value of the reorientation field found in the micromagnetic simulations, with respect to the experiment, can be due to edge effects induced by the limited extension of the simulated cell.[17]

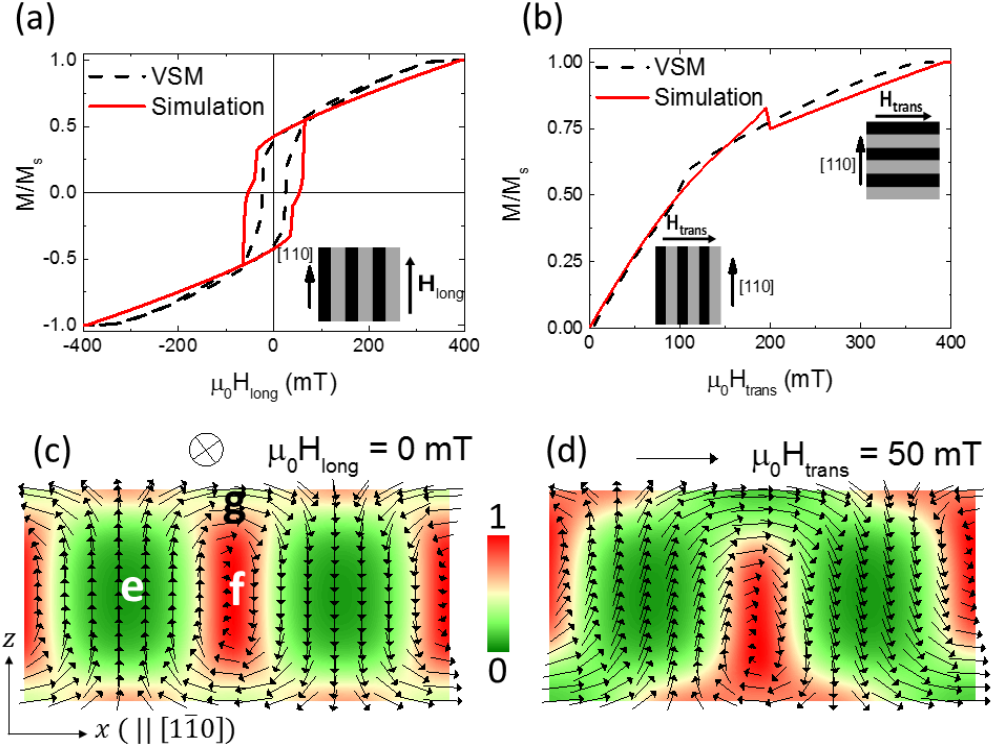

**Fig.2** (Color online) (a) Measured and simulated hysteresis loop of a nitrogen-implanted iron thin film. The measurements were done by VSM along the in-plane [110] direction. (b) Measured and simulated evolution of the magnetization as a function of a magnetic field $\mu_0H_{trans}$ perpendicular to the stripes. (c) Cross-section of the equilibrium magnetization at remanence. The arrows represent the projection of the magnetization in the plane (x,z) while the component $M_y$ is given in a color code from red for high levels to green for low levels. Three kinds of domains can be distinguished: (e) basic domains where the magnetization is essentially perpendicular to the film plane with a slight tilt near the surface, (f) Bloch-type domains in red regions and (g) closure domains near the surface. (d) Equilibrium magnetization over one period of the stripe pattern obtained for a field $\mu_0H_{trans}$ = 50 mT perpendicular to the stripes. An expansion (shrinking) of the closure domains oriented favorably (unfavorably) with the field is observed and the Bloch domains are successively shifted upwards and downwards along the z direction.

*3.3. Dynamic magnetization*

The dynamics of the magnetization of the film has been investigated using two well-established complementary techniques, namely BLS and CPW-FMR. In order to achieve a straightforward comparison of the results of these two techniques, BLS measurements have been performed at normal incidence of light, so that spin waves with zero wavevector were probed. However, it is crucial to remind here that in BLS experiments one probes thermally activated modes that are naturally present in the sample, and the dynamic magnetization component perpendicular to the film surface gives the main contribution to the BLS cross section[38]. On the contrary, in the FMR ones the dynamics is directly excited by a linearly polarized microwave field, and, therefore, only those modes that are compatible with the direction of the exciting field and its spatial uniformity are excited.

*3.3.1. BLS measurements*

Fig. 3(a) and (b) report typical BLS spectra taken at and under an applied field, parallel to the stripes, $\mu_0 H_{long}$= 20 and 160 mT, respectively. Several peaks have been detected, whose frequency monotonically decreases on reducing the external field, as shown by the experimental points in Fig.4 (a). In the same figure, we have reported the results of micromagnetic calculations (color plot), showing a fairly good agreement with the experimental data. Moreover, in Fig.4(b) the spatial profiles of the principal modes, calculated for different values of the external field, are presented. Please note that, as shown in the Fig.S1 of the Supplemental Material, in the case of our relatively complicated domain structure, the results of the simulations strongly depend on the symmetry of the exciting pulse and on the probed film thickness. The best agreement with the experiment, shown in Fig. 4(a), is obtained using an excitation pulse that results from the superposition of several pulses with either even or odd spatial profiles and probing the dynamics of the component of the magnetization perpendicular to the film surface down to a depth of 7.3 nm, i.e. about one tenth of the film thickness, comparable with the penetration depth of light in our sample. This means that only modes with a considerable amplitude at the film surface may exhibit a sizeable cross section.

When the sample is in the saturated state ($\mu_0 H_{long}$=400 mT) four peaks have been detected, corresponding to the perpendicular standing spin wave mode (*n*-PS) characterized by *n* nodes of the magnetization oscillation through the film thickness. When the external field is reduced and the stripes domains

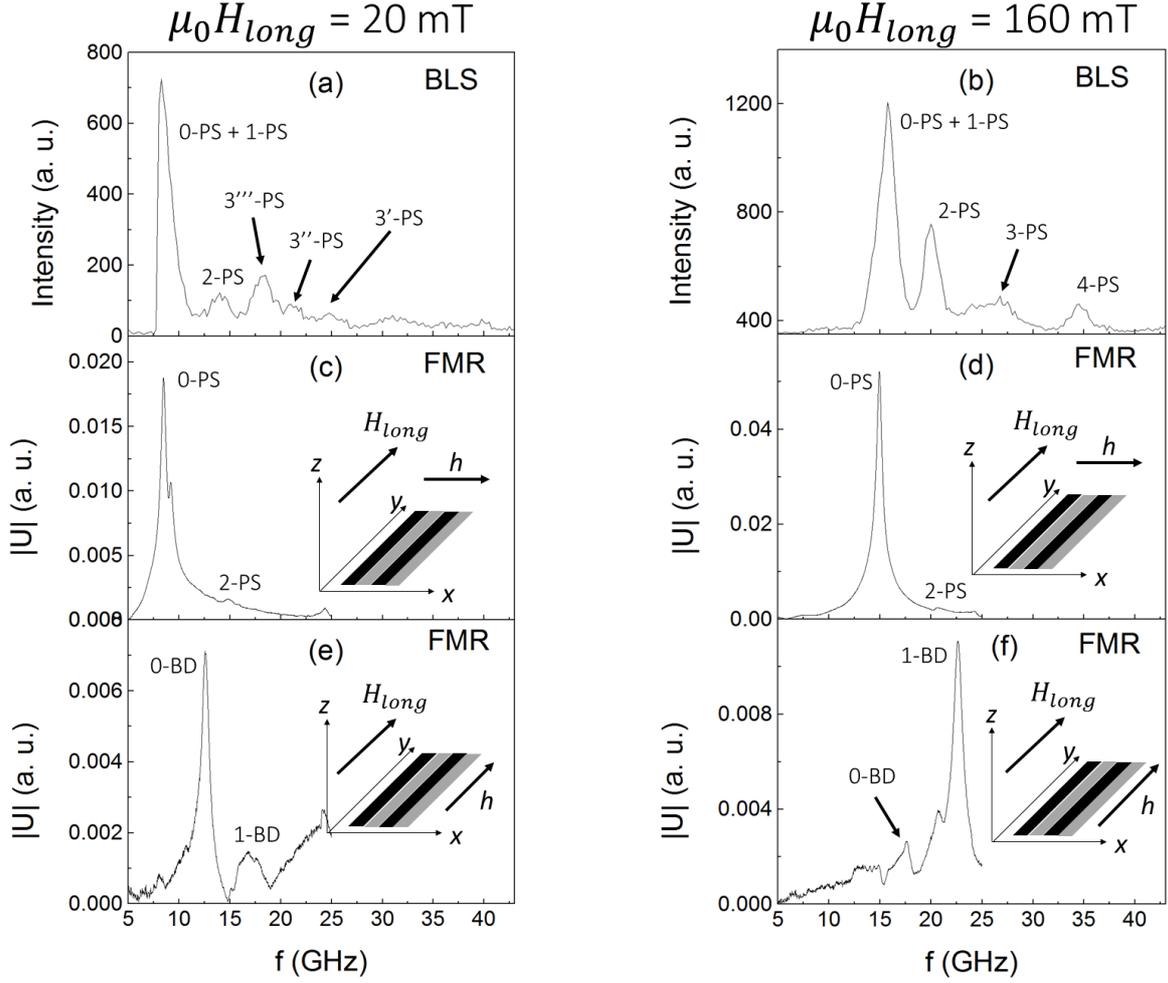

**Fig.3** BLS and CPW-FMR spectra taken at 20 mT (a,c,e) and 160 mT (b,d,f). The static field *H* is parallel to the stripes (along the *y* axis).

appear, these modes localize in different regions of the sample, but still showing *n* nodal planes (although irregular) through the thickness. We therefore keep the label *n*-PS also for the modes in the presence of the stripes. From the simulated spatial profiles, it turns out that the lowest frequency peak seen in Fig. 3 (a) and (b) corresponds to the superposition of the 0-PS (not shown) and 1-PS modes. These two modes are mainly localized in the surface region situated on top of the basic domains and are only different insofar as the precession on the top and bottom surface occurs in-phase or out-of-phase, respectively. The second peak corresponds to the 2-PS mode, characterized by a large spin-precession amplitude both in the Bloch-type domains and on top of the basic domains. At larger frequencies, there is the 3-PS mode that, for magnetic fields lower than about 150 mT, splits in three modes; all of them are localized in the surface region, but the higher frequency mode (3'-

PS) has its largest spin-precession amplitude in the closure domains, whereas the lower frequency mode (3'''-PS) is located on top of the basic domains. The 3''-PS mode, visible only in a limited range of applied field, has a large spin-precession amplitude in the whole surface region, although its cross-section is rather small due to out-of phase precession in adjacent stripes. Finally, the 4-PS mode exhibits a more complicated spatial profile, although one can recognize the presence of four irregular nodal planes in the whole range of fields.

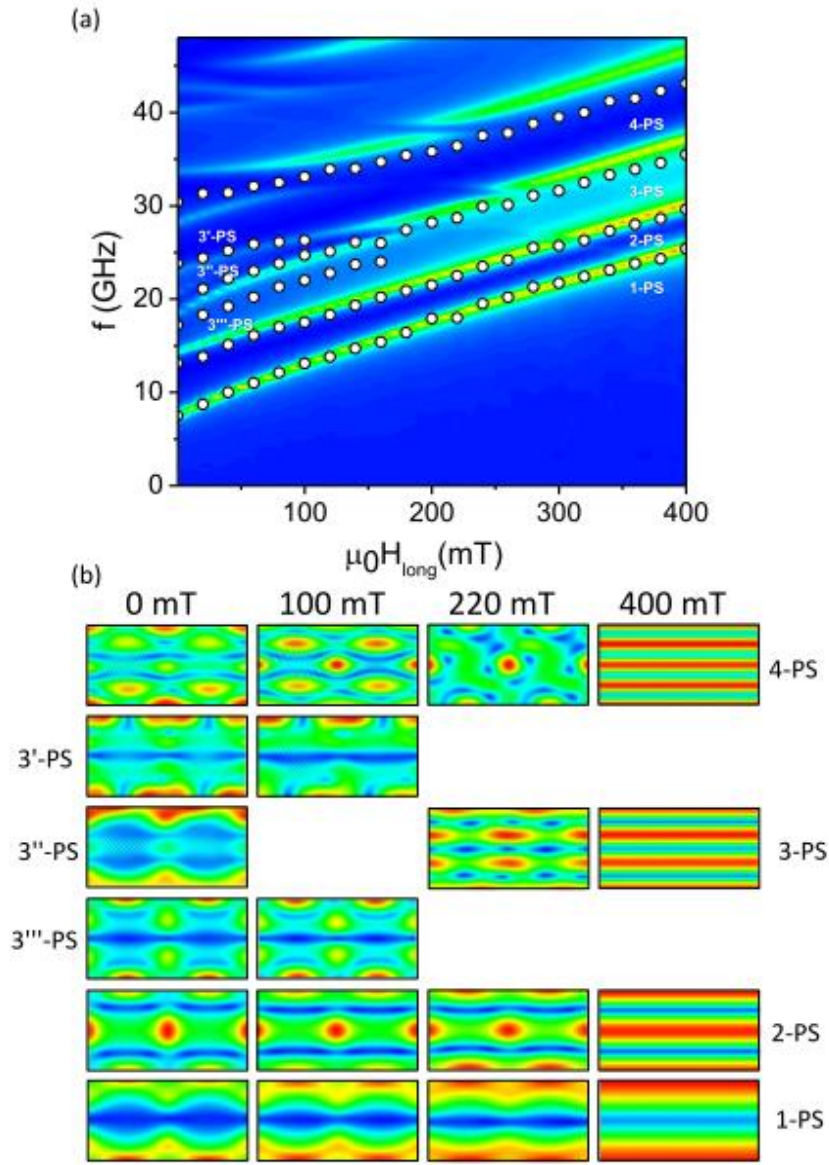

**Fig.4** (Color online) (a) Comparison between the BLS measured (white points) and the calculated (color plot) frequencies as a function of the applied field $H_{long}$. The measurements and simulations are performed decreasing the field from 400 mT to zero. (b) Spatial profile of the modulus of the dynamic magnetization of the principal modes. In all the panels, a color code, from blue for low values to red for high values, has been used.

*3.3.2. CPW-FMR measurements*

Let us now present the results of the CPW-FMR measurements that have been performed in two different geometries. In the first case, both the static field $H_{long}$ and the dynamic one *h(t)* are applied along the *y* axis, as shown in the insets of Fig. 3 (e) and (f), so we will refer to this case as the "*y*-configuration", while in the second case $H_{long}$ is still parallel to *y*, but *h(t)* is applied along the *x* direction, so we will refer to the "*x*-configuration"(Fig.3 (c) and (d)). Therefore, in the *x*-configuration (*y*-configuration) the dynamic exciting field *h(t)* is perpendicular (parallel) to the stripes axis. Figs. 3 (c) and (d) show FMR spectra taken at 20 mT and 160 mT in the *x*-configuration, while the measured and simulated frequencies as a function of the static field $H_{long}$ are reported in Figs. 5 (a) and (b), respectively. In the saturated state ($\mu_0 H_{long}$ =400 mT) only one peak, corresponding to the uniform precession mode (0-PS), was detected. When the field is reduced towards zero and the stripes are formed, this mode takes the same spatial localization of the 1-PS mode, being mainly localized at the surface of the stripes. Moreover, at field around zero a barely visible second peak, corresponding to the 2-PS mode, appears at higher frequency. In contrast with the BLS results shown in the previous section, no higher order PS modes show up in the FMR experiments since the penetration depth of the microwave field is (in contrast with the case of light) much larger than the film thickness, so that odd modes or modes with many nodal planes are not excited in FMR experiments, because the spatial average of their dynamical magnetization is negligible. This is also confirmed by the results of the micromagnetic simulations of Fig. 5 (b), performed exciting the system with a pulse that is uniform along the film thickness.

If we then move to the results of measurements and simulations in the *y*-configuration, Fig. 3 (e) and (f) as well as Fig. 5 (c) and (d), we observe a completely new set of modes, that were not present at all in the previously discussed BLS and FMR data. Looking at the calculated spatial profiles of Fig. 5 (e), one can see that these modes are preferentially localized in the basic domains, where the local magnetization is perpendicular to the surface plane. Therefore, we label them as *n*-Basic Domains (*n*-BD) modes, where the index *n* =0,1,.. indicates modes of increasing frequency. These modes exhibit a surprising non-monotonic behavior as a function of the applied field, which can be explained on the basis of the evolution of both their spatial profiles and the stripe configuration. First of all, when the film is in the saturated state ($\mu_0 H_{long} \geq 400$ mT) no BD mode exists, because the sample magnetization is parallel to the surface plane. On decreasing the external field from saturation, however, the magnetization in the center of the adjacent stripes starts to rotate up- or downwards with respect to the film surface and the 0-BD mode appears at very low-frequency. On further reducing the applied field, its frequency initially increases (reflecting the gradual rotation of the magnetization) and then decreases when the mode takes a surface localization. One can also see that higher order BD modes appear at specific values of the external field and all of them are localized, when they start to exist, in the inner part of the stripes. However, the non-monotonic frequency behavior is not observed for the 3- and 4-BD modes, because they appear at relatively low field and remain always localized in the basic domains. As a general consideration on the frequency evolution of these n-BD modes, one can notice the analogy to the case of the resonant frequency of ultrathin films characterized by PMA, where a non-monotonic frequency behavior is seen, in coincidence with the re-orientation of the magnetization from parallel to perpendicular to the film surface.[39]

Finally, let us notice that it appears particularly interesting the selective and mutually exclusive presence of either the PS or the BD set of modes in the *x*- or *y*-configuration of FMR measurements and simulations. This is also due to the phase correlation of the dynamical magnetization components in adjacent stripes. In particular, our simulation show (see Fig. S2 in the Supplemental Material) that the spatial average of either the *y*- or *x*-component of the dynamic magnetization is zero for the PS or BD modes, respectively, because they oscillate out-of-phase in adjacent stripes.[14] Therefore, only the PS or the BD family of modes is selectively present in FMR experiments performed in the *x*- or *y*-configuration. As for the absence of BD modes in the BLS results of the previous section, one should again consider the limited penetration depth of light within the metallic film (a few nanometers) and the major role of the dynamic component $m_z$ to the BLS cross-section. Looking at the spatial profiles of the different components of the magnetization in Fig. S2 of the Supplemental Material, it appears that $m_z$ oscillate out-of-phase in adjacent stripes, so that the corresponding BLS cross section is negligible.

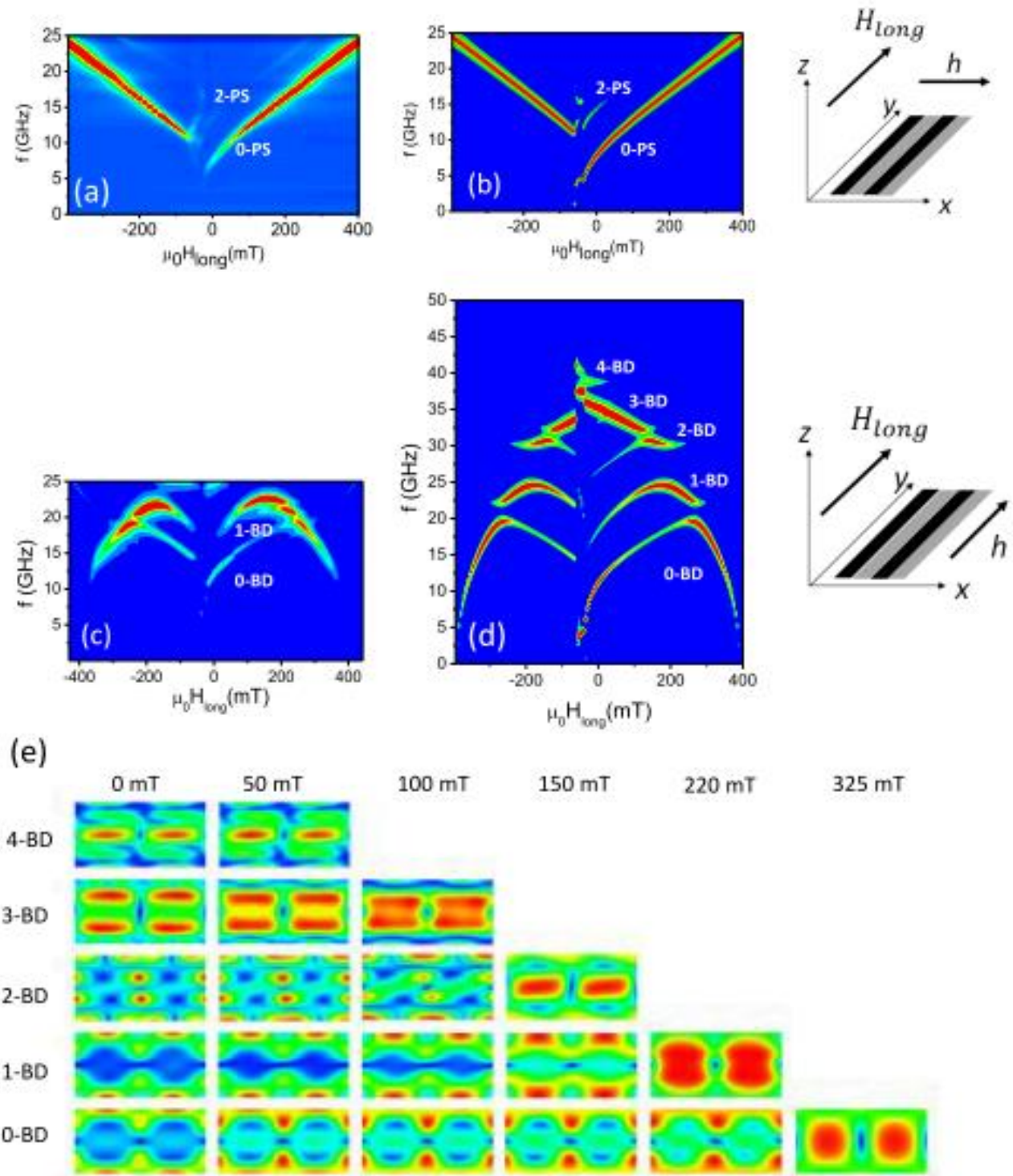

**Fig.5** (Color online) Measured (a,c) and simulated (b,d) FMR spectra field in *x*-configuration (a,b) and *y*-configuration (c,d). The measurements and simulations are performed decreasing the field from 400 mT to zero and then increasing the field in the negative direction up to -400 mT. Due to the instrument limitation, the maximum frequency accessible in FMR experiments was 25 GHz. (e) Spatial profile of the modulus of the dynamic magnetization of the principal modes in *y*-configuration. In all the panels, a color code, from blue for low values to red for high values, has been used.

### 3.3.3. Rotation of the stripes

Let us now focus on the dynamic properties of the stripes domains when an in-plane field is applied along the transverse direction (with respect to the stripes axis). First, the stripes domains have been prepared along the [110] direction, then a field $\mu_0H_{trans}$ of increasing intensity is applied along the in-plane transverse direction, similar to the experiment of Fig. 2(b). Fig.6 shows the comparison between the spin-wave frequencies measured by BLS and the micromagnetic simulations. We found a quite good agreement, excluding the field range between about 50 and 200 mT, because, as already mentioned in Sect. III B, in the experiments the stripes start to rotate at 100 mT, while a double value is found in the micromagnetic simulations. At zero field the observed modes are the same already found in Fig.4 (b), so they are labelled 1-PS, 2-PS, 3"-PS, and 3'-PS. On increasing the field intensity, however, the stripe configuration is deformed and also the mode profiles evolve, until at about 100 mT (200 mT in the simulations) the stripes abruptly rotate and there is a discontinuity in the modes frequencies caused by the stripes rotation. After the reorientation, the external field is parallel to the stripes axis and the usual *n*-PS modes are restored. Let us focus now on the behavior of the 1-PS mode that, at zero field, has the maximum amplitude in the surface region situated on top of the basic domains, as seen in Fig. 6(b). When the stripes domain structure is distorted, under the influence $H_{trans}$, the mode localizes in the closure domains where the magnetization is almost antiparallel with respect to $H_{trans}$. This is illustrated in the bottom panels of Fig. 6. As a consequence, its frequency remains almost constant, while its intensity gradually decreases, until this mode suddenly disappears when the stripes rotate along $H_{trans}$. In the case of the higher order PS modes, the frequency dependence is even more complicated, but one can follow a relatively continuous evolution of their frequency until they attain the ordinary perpendicularly standing modes of the saturated film.

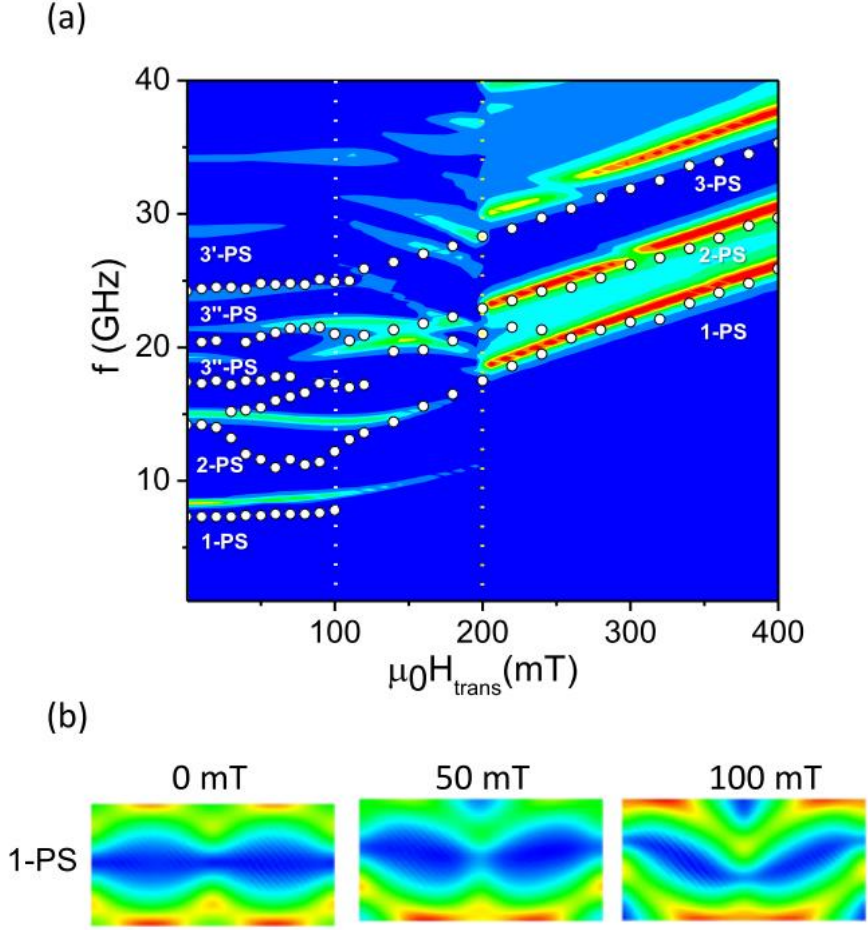

**Fig.6** (Color online) (a) Comparison between the measured (white points) and the calculated (color plot) frequencies measured as a function of the $\mu_0 H_{trans}$ field, that is swept from zero to 400 mT, along the direction perpendicular to the stripes. (b) Spatial profile of the modulus of the dynamic magnetization of the 1-PS mode as a function of the $\mu_0 H_{trans}$ intensity. In all the panels, a color code, from blue for low values to red for high values, has been used.

The rotation of the stripes under the action of $\mu_0 H_{trans}$ has been also investigated by CPW-FMR measurements and simulations, performed in the two possible configurations shown in Fig.7 (a-b) and (c-d), respectively. In the former configuration, the dynamical probing field is orthogonal to the initial stripe direction, so at remanence one can observe the 0-PS and 2-PS modes, as expected in analogy with Fig. 5(a-b). On increasing $\mu_0 H_{trans}$ the former mode exhibits a frequency evolution in good agreement with the 1-PS mode observed in the BLS measurements, since when the stripes are present these two modes are characterized by the same spatial localization and almost degenerate frequencies. Remarkably, for $\mu_0 H_{trans}$ intensity larger than about 100 mT (200 mT in the simula-tions), the stripes rotate along the field direction, and, as a consequence, BD modes suddenly appear in the experiment, while PS modes disappear since they do not couple anymore to the exciting field. A similar, but opposite, transition from one family of modes to the other is also observed if one performs CPW-FMR experiments or simulations using a probing dynamic field that is initially parallel to the stripe direction, as reported in Fig. 7(c-d). In such a case, at remanence and for small applied field one sees low-order modes belonging to the BD family, while the 0-PS mode is recovered after the reorientation of the stripes. One may thus conclude that while BLS is always sensitive to PS-like modes independently of the orientation of the stripes, in the FMR experiments one switches from PS-like to BD-like modes or vice versa, as a consequence of the stripe reorientation.

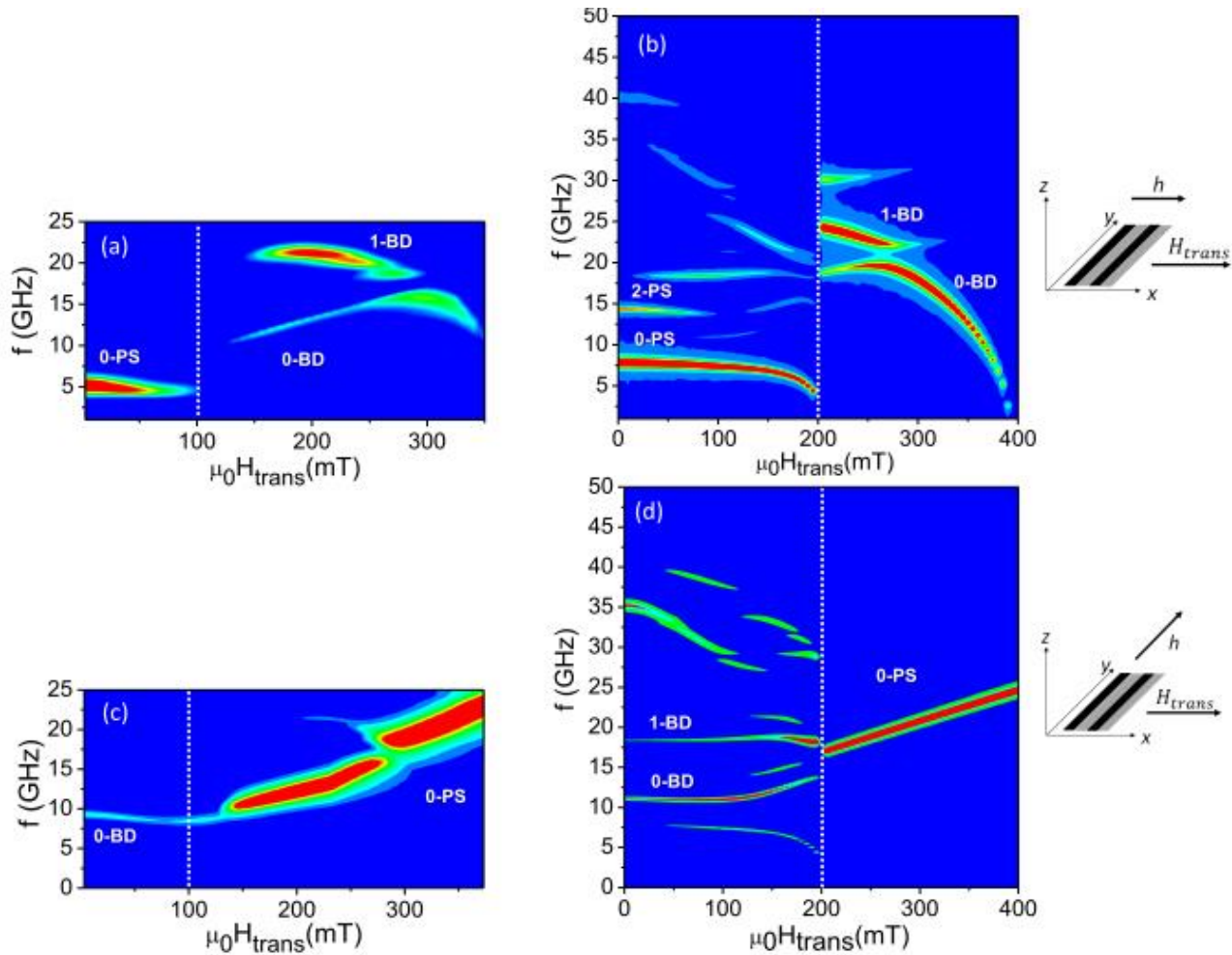

**Fig.7** (Color online) Measured (a,c) and simulated (b,d) FMR spectra field as a function of the intensity of an in-plane magnetic field $\mu_0 H_{trans}$ applied in the direction perpendicular to the stripes axis. In all the panels, a color code, from blue for low values to red for high values, has been used.

## 4. Conclusion

In this work we exploited BLS and CPW-FMR to investigate the magnetization dynamics of a weak stripes pattern in an α'-FeN film. Thanks to the complementarity of the two techniques and the comparison with micromagnetic simulations, the main features of the eigenmodes spectrum, at zero wavevector, have been explained. In particular, we observed a first set of modes, preferentially localized at the surface of the stripes domains, that could be selectively revealed by BLS measurements. These modes continuously transform into the usual perpendicular standing modes of the saturated film for sufficiently large applied field. A second family of modes was instead observed in CPW-FMR with microwave field orthogonal to the stripes direction. These modes are localized in the basic domains at the inner region of the stripes where the local magnetization is perpendicular to the surface plane and disappear when the film is uniformly magnetized. The evolution of both the frequency and the spatial profile of the two families of modes has been correlated to the evolution of the stripe domain structure as a function of the intensity of an external magnetic field, applied along the direction either parallel or perpendicular to the stripes axis.

We are confident that the present study will be a milestone also towards the extension of the study to spin waves with finite wavevectors, in view of the possible exploitation of FeN films as natural magnonic crystals. In fact, the existence of stripe domains should induce the occurrence of alternating allowed and forbidden band gaps in the dispersion curves of spin waves propagating across the stripes, making FeN films attractive for the emerging field of magnonics.

## Acknowledgments

We gratefully thank Marie Vannini and Frédéric Molina at Rohde & Schwarz France S.A.S. for freely providing us a R&S R ZVA40 Vector Network Analyzer. We also thank Mathieu Bernard (INSP) for his assistance in building the coplanar waveguide mounting structure, Emmanuel Lhuillier and Laura Thevenard (INSP) for their help in the resistivity measurements and Paola Atkinson for preparing GaAs substrates. This work was supported by the Agence Nationale de la Recherche (France) under contract No. ANR-13-JS04-0001-01 (SPINSAW) and the MATINNOV industrial chair of the Agence Nationale de la Recherche, resulting from the collaboration between the University of Versailles Saint-Quentin-en-Yvelines and VALEO.